\documentstyle[aps,prb]{revtex}
\newcommand{\be}{\begin{equation}}
\newcommand{\ee}{\end{equation}}
\newcommand{\calp}{\mbox{$\mathcal P$}}
\begin{document}
\draft
\title{The Penalty Method for Random Walks with Uncertain Energies}
\author{ D. M. Ceperley and M. Dewing\\
Department of Physics and NCSA\\
University of Illinois at Urbana-Champaign, Urbana, IL 61801}
\date{\today}
\maketitle
\begin{abstract}
We generalize the Metropolis {\it et al.} random walk algorithm to
the situation where the energy is noisy and can only be estimated.
Two possible applications are for long range potentials and for
mixed quantum-classical simulations. If the noise is normally
distributed we are able to modify the acceptance probability by
applying a penalty to the energy difference and thereby achieve
exact sampling even with very strong noise.  When one has to
estimate the variance we have an approximate formula, good in the
limit of large number of independent estimates. We argue that the
penalty method is nearly optimal.  We also adapt an existing
method by Kennedy and Kuti  and compare to the penalty method on a
one dimensional double well.
\end{abstract}
\pacs{ 05.10.Ln, 02.70.Lq, 07.05.Tp}

\section{Introduction}
As Metropolis {\it et al.} showed in 1953\cite{mrt}, Markov random
walks can be used to sample the Boltzmann distribution thereby
calculate thermodynamic properties of classical many-body systems.
The algorithm they introduced is one of the most important and
pervasive numerical algorithms used on computers because it is a
general method of sampling arbitrary highly-dimensional
probability distributions. Since then many extensions have been
developed\cite{MCreview}. In addition to the sampling of classical
systems, many Quantum Monte Carlo algorithms such as Path Integral
Monte Carlo\cite{rmp}, variational Monte Carlo\cite{cck} and
Lattice Gauge Monte Carlo use a generalization of the random walk
algorithm.

In a Markov process, one changes the state of the system $\{s \}$
randomly according to a fixed {\it transition rule},
${\calp}(s\rightarrow s')$, thus generating a random walk through
state space, $\{s_0,s_1,s_2 \ldots \}$. The transition
probabilities often satisfy the {\it detailed balance} property (a
sufficient but not necessary condition). This means that the
transition rate from $s$ to $s'$ equals the reverse rate:
\begin{equation}
\pi (s) {\calp}(s\rightarrow s')=\pi (s') {\calp}(s'\rightarrow s).
\label{db}
\end{equation}
Here $\pi(s)$ is the desired equilibrium distribution which we
take for simplicity to be the classical Boltzmann distribution:
$\pi(s) \propto \exp(- V(s)/(k_B T))$ where $T$ is the temperature
and $V(s)$ is the energy. If the pair of functions $ \{ \pi (s),
{\calp}(s\rightarrow s') \} $ satisfy detailed balance and if
${\calp}(s\rightarrow s')$ is ergodic, then the random walk will
eventually converge to $\pi$. For more details see Refs. [\onlinecite{hh,wk}].

In the particular method introduced by Metropolis one ensures that
the transition rule satisfies detailed balance by splitting it
into an ``a priori'' {\it sampling distribution} $T(s\rightarrow
s')$ (a probability distribution that can be directly sampled such
as a uniform distribution about the current position) and an {\it
acceptance probability} $a(s\rightarrow s')$ with $0 \le a \le 1$.
The overall transition rate is:
\begin{equation}
{\calp}(s\rightarrow s') = T(s\rightarrow s') a(s\rightarrow s').
\end{equation}
Metropolis et al. \cite{mrt} made the choice for the acceptance
probability:
\begin{equation} 
a_M (s\rightarrow s') = \min \left[ 1 , q (s'\rightarrow s) \right], 
\label{acc}
\end{equation}
where
\begin{equation}
q(s\rightarrow s') =  \frac{\pi (s') T (s'\rightarrow s)  } {\pi
(s) T (s\rightarrow s') } = \exp(-(V(s')-V(s))/(k_BT)).
\label{qdef}
\end{equation}
Here we are assuming for the sake of simplicity that $T
(s'\rightarrow s) = T (s\rightarrow s')$. The random walk does not
simply proceed downhill; thermal fluctuations can drive it uphill.
Moves that lower the potential energy are always accepted but
moves that raise the potential energy are often accepted if the
energy cost (relative to $k_B T= 1/\beta$) is small. Since
asymptotic convergence can be guaranteed, the main issue is
whether configuration space is explored thoroughly in a reasonable
amount of computer time.

What we consider in this article is the common situation where the
energy, $V(s)$ needed to accept or reject moves, is itself
uncertain. This can come about because of two related situations:
\begin{itemize}
\item
The energy may be expressed as an integral: $V(s) = \int dx
v(x,s)$. If the integral has many dimensions, one might need to
perform the integral with another subsidiary  Monte Carlo
calculation.
\item The energy may be expressed as a finite sum: $ V(s) =
\sum_{k=1}^N e_k(s)$ where is $N$ is large enough that performing
the summation slows the calculation.  It might be desirable for
the sake of efficiency to sample only a few terms in the sum.
\end{itemize}

\subsection{Mixed Quantum-Classical Simulation}

First, consider the typical system in condensed matter physics and
chemistry, composed of a number of classical nuclei and quantum
electrons.  In many cases the electrons can be assumed to be in
their ground state and to follow the nuclei adiabatically. To
perform a simulation of this system, we need to accept or reject
the nuclear moves based on the Born-Oppenheimer potential energy
$V_{BO} (s)$, defined as the eigenvalue of the electronic
Schr\^{o}dinger equation with the nuclei fixed at position $s$. In
most applications, this potential is approximated by a
semi-empirical potential typically involving sums over pair of
particles. More recently, in the Car-Parrinello molecular dynamics
method\cite{cp}, one performs a molecular dynamics simulation of
the ions simultaneous with a solution of the electronic quantum
wave equation. To be feasible one uses a mean field approximation
to the full many-body Schr\"{o}dinger equation using the local
density functional approximation to density functional theory or a
variant. Others have proposed coupling a nuclear Monte Carlo
random walk to an LDA calculation\cite{Kohn}. Although mean-field
methods such as LDA are among the most accurate methods fast
enough to be useful for large systems, they also have known
deficiencies\cite{mitas}.

We would like to use a quantum Monte Carlo (QMC) simulation to
calculate $V_{BO}(s)$ during the midst of a classical MC
simulation (CMC) \cite{footnote1}.
QMC methods, though
not yet rigorous because of the fermion sign problem, are the most
accurate methods useful for hundreds of electrons. But QMC
simulation will only give an estimate of $V_{BO}(s)$  with some
statistical uncertainty. It is very time consuming to reduce the
error to a negligible level. We would like to take into account
the statistical error without having to reduce it to zero.

Note that we do not wish the new Monte Carlo procedure to
introduce uncontrolled approximations because the goal of coupling
the CMC and QMC is a robust, accurate method. We need to control
systematic errors. It has been noticed by Doll and Freeman
\cite{doll}, after studying a simple example, that CMC
is robust with respect to noise but recommend using small noise
levels and small step sizes to minimize the systematic errors.
However, this can degrade the overall efficiency. If we can
tolerate higher noise levels without introducing systematic
errors, the overall computer algorithm will run faster and more
challenging physical systems can be investigated, {\it e.g.} more
electrons and lower temperatures

\subsection{Long-range potentials }

In CMC with a pair potential, to compute the change in energy when
particle $k$ is moved to position ${\bf r}_k'$, one needs to
compute the sum 
\be
\label{sumpot}
\Delta V ({\bf r}_k')= \sum_{j=1}^N [v(r_{kj}')-v(r_{kj})]. 
\ee
This is referred to as an order $N^2$
algorithm since the computer effort to move all particles once is
proportional to $N^2$. If the interaction has a finite range,
neighbor tables\cite{allentild} will reduce this complexity to
order $N$. However charged systems with Coulomb interactions are
not amenable to this treatment. Usually the Ewald image method is
used to handle the long-range potentials with a complexity
\cite{NatCep} of order $N^{3/2}$. The fast multipole
method\cite{greengard}, which scales as $N$ for the Coulomb
interaction is not applicable to Monte Carlo since that method
computes the total energy or force and in MC we need the change in
potential as a single particle is moved.

The challenge is to come up with an order $N$ Monte Carlo method
for charged systems. In the Ewald method, the potential is split
into a short-range part and a long-range part: \be v(r) = v_s(r)
+v_l (r).\ee The short ranged part is a finite ranged and can be
handled with neighbor tables, the long range part is usually
expanded in a Fourier series, at least in periodic boundary
conditions and is bounded and slowly varying. We suggest that it
is possible to estimate the value of $v_l (r)$ by sampling either
particles at random, or terms in its Fourier expansion. The
question that arises is how to compensate for the noise of the
estimate in $\Delta V$.


In both of these examples one could simply ignore the effect of
fluctuations in the estimate of $\Delta V(s)$. If the errors are
small then clearly the sampled distribution will be changed only a
little. If the acceptance ratio as a function of $\Delta V(s)$
were a linear function there would be no bias, but because it is
non-linear, fluctuations will bias the asymptotic distribution. In
this paper we will make a conceptually simple generalization of
Metropolis algorithm, by adjusting the acceptance ratio formula so
that the transition probabilities are unaffected by the
fluctuations in the estimate of $\Delta V(s)$. We end up with a
completely rigorous formula in the sense that if one averages long
enough, one will get the exact distribution, even if the noise
level is large. The only assumption is that the individual energy
estimates are independently sampled from a normal distribution
whose mean value is $\Delta V(s)$. One complication is that the
estimates of the variance of $\Delta V(s)$ are also needed. We
show how to treat that case as well.

Kennedy, Kuti and Bhanot\cite{kk,bhanot} introduced an algorithm
with many of the same aims as the present work but for
computations in lattice gauge theory. We will describe their
method and compare it to the new method later in the paper.

\section{Detailed balance with uncertainty.}

From the two examples discussed above, let us suppose that when a
move from $s$ to $s'$ is made, an estimate of the difference in
energy is available, which we denote $\delta(s \rightarrow s')$.
(We often take units with $k_BT=1$ hereafter.) By $V(s)$ we mean the
true potential energy. Let $a(s \rightarrow s')$ be a modified
acceptance probability; we assume that it depends only on the
estimate $\delta$ of the energy difference.  Let $P( \delta ; s
\rightarrow s' )d\delta$ be the probability for obtaining a value
$\delta$. Then the average acceptance ratio from $s$ to $s'$:
\be
A(s \rightarrow s') =\int_{-\infty}^{\infty} d\delta P(\delta; s
\rightarrow s' ) a(\delta). \label{defA} \ee The detailed balance
equation is:
 \be
 e^{-V(s)/k_BT}T(s \rightarrow s')A( s \rightarrow s' ) = 
 e^{-V(s')/k_BT}T(s' \rightarrow s)A( s' \rightarrow s ) 
\ee
Defining:
\be
\Delta(s \rightarrow s') = [V(s')-V(s)]/k_BT 
 -\ln[T(s' \rightarrow s)/T(s \rightarrow s')] 
\ee 
we can rewrite the detailed balance
equation as:
\be
A(s \rightarrow s ') = e^{-\Delta} A( s' \rightarrow s ).
\label{MDB} \ee If the process to estimate $\delta$ is symmetric
in $s$ and $s'$ then $ P(\delta; s' \rightarrow s) = P(-\delta;s
\rightarrow s')$. Then detailed balance requires:
\be
\int_{-\infty}^{\infty} d\delta P(\delta; s \rightarrow s' ) [ a
(\delta) - e^{-\Delta} a(-\delta) ] = 0. \label{inteq} \ee In
addition, we must have that $0 \leq a(\delta ) \leq 1$ since $a$
is a probability \cite{footnote2}.

The difficulty in using these formulas is that during the MC
random walk, we do not know either $P(\delta; s \rightarrow s' )$
or $\Delta$. Hence we must find a function $a(\delta)$ which
satisfies Eq. (\ref{inteq}) for all $P(\delta)$ and $\Delta $.

To make progress we assume a particular form for $P(\delta; s
\rightarrow s' )$.  In many interesting cases, the noise of the
energy difference will be normally distributed. In fact the
central limit theorem guarantees that the probability distribution
of $\delta$ will approach a normal distribution if the variance of
the energy difference exists and one averages long enough. 
Given that $\langle \delta \rangle = \Delta$,
the probability of getting a particular value of $\delta$ is:
\begin{equation}
P(\delta) = (2 \sigma^2 \pi)^{-1/2}
\exp(-(\delta-\Delta)^2/(2\sigma^2)). \label{normal}
\end{equation}
In this section only, we will assume that we know the value of
$\sigma$, that only $\Delta$ is unknown. We will discuss relaxing
this assumption in Sec. \ref{sigmasec}.

In the case of a normal distribution with known variance $\sigma$
we have found a very simple exact solution to Eq. (\ref{inteq}):
\begin{equation}
a_P(\delta; \sigma ) =  \min(1, \exp(-\delta-\sigma^2/2) )
\label{1psol}
\end{equation}
The uncertainty in the action just causes a reduction in the
acceptance probability by an amount $\exp(-\sigma^2/2)$ for 
$\delta > -\sigma^2/2$. We refer
to the quantity $u= \sigma^2/2$ as the noise {\it penalty}. 
Clearly, the formula
reverts to the usual Metropolis formula when the noise vanishes.

To prove Eq. (\ref{1psol}) satisfies Eq. (\ref{MDB}), one does the
integrals in Eq. (\ref{defA}) to obtain:
\begin{equation}
A( \Delta ) = \frac{1}{2} [e^{-\Delta} \text{erfc}(c(\sigma^2/2-\Delta))+
 \text{erfc} (c(\sigma^2/2+\Delta))]
\end{equation}
where $\text{erfc}(z)$ is the complimentary error function and
$c=1/\sqrt{2\sigma^2}$.

Below we apply Eq. (\ref{1psol}) to several simple problems and
find that it indeed gives exact answers to statistical precision.
The remainder of the paper concerns considerations of efficiency,
a comparison to other methods and the more difficult problem of
estimating $\sigma$.

\section{Optimality}

The chief motivation for studying the effect of noise on a Markov
process is for reasons of efficiency. If computer time were not an
issue, we could average enough to reduce the noise level to an
insignificant level.  In this section we are concerned with the 
question of how to optimize the acceptance formula and the noise
level. 

\subsection{Acceptance ratio}

We first propose a measure of optimality of an acceptance
formula and relate that to a linear programming problem.
It is clear that Eq.(\ref{inteq}) can have multiple solutions; its
solution set is convex. For example, if $a(\delta)$ is a solution
then so is $\lambda a(\delta)$ for $0 < \lambda <1$. Even in the
noise-less case, several acceptance formulas have been suggested
in the literature\cite{wood,barker}. To choose between various
solutions we now discuss the efficiency of the Markov process,
namely the computer time needed to calculate a property to a given
accuracy. It is a difficult problem\cite{mira} to determine the
efficiency of a Markov chain but Peskun\cite{peskun73} has shown
that given two acceptance rules, $a_1(x)$ and $a_2(x)$, if $a_1
(\Delta) \ge a_2(\Delta)$ for all $ \Delta \neq 0 $, then every
property will be computed with a lower variance using rule 1
versus rule 2. Hence the most efficient simulation will have the
maximum value of $\lambda$. Very roughly what Peskun has shown is
that it is always better to accept moves, other considerations
being equal.

We propose to call an {\it optimal} acceptance formula, one where
the average probability of moving is as large as possible. Let
$W(\delta)d\delta$ be the probability density of attempting a move
with a change in action $\delta$, ( $W(\delta) \geq 0$.) In our
definition an ``optimal'' formula will maximize:
\begin{equation}
\xi =\int_{-\infty}^{\infty} d \delta W( \delta )\left( a( \delta)-
a_M( \delta)\right). \label{optimum}
\end{equation}
It is likely that the optimal functions are, to a large part,
independent of $W$ and so we set $W(x)=1$. We subtracted $a_M(x)$,
 the Metropolis formula, so the integral would be convergent.
Note that for the solution for a normal distribution $a_P(\delta)$
we have: $\xi_P=-\sigma^2/2$.

In the noise-less case one can easily show\cite{peskun73} that the
Metropolis formula is optimal. Without uncertainty, Eq.
(\ref{MDB}) only couples values with the same $|\delta|$:
$a(\delta)=e^{-\delta}a(-\delta)$. For each $\delta
> 0 $, one needs to maximize: $W(\delta) a(\delta) + W(-\delta) a(
-\delta)$. This and the constraint  $0 \leq a(x)\leq 1 $ leads to
the solution $a(\delta)=1$ if $\delta \leq 0$.

We conjecture that the formula Eq. (\ref{1psol}) is nearly
optimal; one argument is based on an analysis of the large and
small $\delta$ limits: the other is numerical. First, consider
moves which are definitively uphill or downhill $ \delta^2 \gg
\sigma^2$. We expect downhill moves will always be accepted for 
an optimal function, so $A(\Delta)=1$; this is its maximum
value. 
Then from Eq. (\ref{MDB}) $A(
\Delta) =e^{-\Delta}$ for $\Delta \gg \sigma$. Now we must invert
Eq. (\ref{defA}). The unique continuous solution is
$a(\delta)=\exp(-\delta-\sigma^2/2)$ for $\delta \gg \sigma$.
Hence, in the region $|\delta| \gg \sigma$ the solution is optimal
in the class of continuous functions \cite{footnote3}.

Another approach to finding the optimal solution to Eq.
(\ref{inteq}) is numerical. We wish to maximize Eq. (\ref{optimum})
subject to equality constraints and the inequality constraints
that $ a(\delta)$ be a probability. This is an infinite
dimensional linear programming (LP) problem, a well-studied
problem in optimization theory for which there exist methods to
determine the globally optimal solution. To find such a solution,
we represent $a(\delta)$ on a finite basis. We used a uniform grid
in the range $-y$ to $y$ and assumed that outside the range
$a(\delta)$ had the asymptotic form derived above. 
The discrete version of Eq. (\ref{defA}) is
$A_j = \sum_i K_{ij} a_i + c_j$ where
$c_j$ represents the contribution coming from $|\delta|> y$ and
$K_{ij} = P(\delta_i;\Delta_j)$ for the simplest quadrature.
The problem is to find a solution maximizing $\sum_i a_i$ subject to
the inequalities: $0 \leq a \leq 1$ and the equalities: \be\sum_i
[K_{i,j}-e^{-x_j}K_{i,-j}] a_i =e^{-x_j}c_{-j}- c_j.\ee

Fig. \ref{LP} shows the LP solution, for $\sigma =1$ compared with
$a_P(\delta)$. Note that it is not a continuous function, but for
the most part consists of regions with $a_i = 1$ alternating with
regions with $a_i=0$. The LP solution is a very accurate solution
to the problem posed, with errors of less than 10$^{-5}$. The
discontinuous nature of the LP solution is to be expected since
the solution must lie on the vertices of the feasible region,
determined by the equalities and inequalities. To obtain the
solution to this difficult ill-conditioned problem, we discretized the 
values of $\delta$ on a grid with
spacing 0.01. However we only demanded that Eq. (\ref{MDB}) be
satisfied on a grid $\Delta$ with a spacing of 0.2. This implies
that there were 40 times as many degrees of freedom as equality
constraints and thus most variables were free to reach the
extreme values of 0 and 1.

The optimal LP function has a slightly larger value of $\xi$,
roughly about $\xi_{LP} \approx -0.45 \sigma^2$ versus the value
for $a_p$ of $\xi_P = -0.5 \sigma^2$. As far as we can
determine, the LP solutions survive in the limit $dr \rightarrow
0$ and are slightly more optimal than $a_P$. However, given the
inconvenience of determining and programming the LP solutions, and
the very limited improvement in $\xi$, we see little
reason \cite{footnote4}
to prefer such solutions. When we added a factor to
penalize discontinuities in $a(\delta)$ to the objective function
proportional to $\sum_i (a_i-a_{i-1})^2$ (this makes it a
quadratic programming problem) then the solution converged to
$a_P(\delta)$.

\subsection{Noise level}
\label{noiseopt}

Now let us consider how to optimize the noise level $\sigma$.
An energy difference with a large noise level can be computed quickly,
but because of the penalty in Eq. (\ref{1psol}) it has a low acceptance ratio,
reducing the overall efficiency of the simulation.
We should pick $\sigma$ to minimize the variance
of some property with the total computer time fixed. The computer
time can be written as $T = m (n t +t_0)$ where $t$ is the time
for an elementary evaluation of a given energy difference, $n$ is
the number of evaluations of $\delta$ before an acceptance is
tried, $m$ is the total number of steps of the random walk and
$t_0$ is the CPU time in the noise-less part of the code. But the
error in any property converges as $\epsilon = c (\sigma)
m^{-1/2}$ where $c$ is some function of $\sigma$ and the noise
level converges as $\sigma = d n^{-1/2}$ where $d$ is some constant. 
Eliminating the variables $m$ and $n$, we write the MC inefficiency:
\begin{equation}
\label{effeq}
\zeta^{-1} =T \epsilon^2= t_0 c(\sigma)^2 \left[ f\sigma^{-2} +1
\right].
\end{equation}
Here $f=d^2t/t_0$, the relative noise parameter, is the CPU time
needed to reduce the variance of the energy difference relative to
the CPU time used in the noise-less part of the code: for $f \ll
1$ noise is unimportant, for $f \gg 1$ computation of the noisy energy
difference dominates the computer time.

To demonstrate how important this optimization step is, we consider
a one dimensional double well with a potential given by: \be k_B T
V(s) = a_1 s^2 + a_2 s^4 .\ee We picked parameters such the two
minima are at $s=\pm 4$ and the height of the central peak is 
$\pi(0)/\pi(4) = 0.1$, which 
corresponds to
$a_1=-0.288 $ and $a_2=0.009$. We used a uniform transition
probability ($T(s\rightarrow s')$) with a maximum move step of
0.5. This means overcoming the barrier requires multiple steps,
typical of an application which has a probability density with
several competing minima. To measure the efficiency, we computed
the error on the average value of $\langle s^k\rangle$
on Markov chains
with $10^7$ steps. We examined values of noise in the range $0 \leq \sigma
\leq 6$. We also calculated the density and compared to the exact
values obtained by deterministic integration. Shown in Fig. \ref{accratio} is
the acceptance ratio versus $\sigma$. We see that it decreases to
zero rapidly at large noise levels. The dotted line ($\propto
\exp[-\sigma^2/8]$) is the asymptotic form for large $\sigma$.

Fig. \ref{doublewell} shows an example of the density obtained when the noise
in the energy was $\sigma=2$. It is seen that ignoring the noise
leads to a much smoother density than the exact result. Using the
acceptance formula $a_P (\delta)$ we recover the exact result
within statistical errors.

Figure \ref{eff} shows the inefficiency (relative to its value when the
noise is switched off) versus $\sigma$ and $f$. In general, as the
difficulty of reducing the noise (as measured by $f$) increases, the
calculation becomes less efficient, and the optimal value of
$\sigma$ increases. The two panels show the efficiency of computing
$\langle s \rangle$ and $\langle s^2 \rangle$; the behavior of the
error is quite different for even and odd moments of $s$ because
the error in the first moment is sensitive to the rate at which
the walk passes over the barrier, while the second is not. The
flat behavior at large noise level of the first moment occurs
because the noise actually helps passage over the barrier: for
$f>3$ a finite optimal value of $\sigma$ ceases to exist.

On this example, we find that
$c(\sigma)\propto\exp(\alpha\sigma^2)$ with $\alpha \approx 0.09$ for even
moments and $\alpha \approx 0.025$ for odd moments. With this
assumption the optimal value of the noise level equals: \be
\sigma^{*2}= (f/2)[ \sqrt{1+2/(f\alpha)}-1].\ee Although this
formula is approximate (because of the assumption on $c(\sigma)$)
it does give reasonable values for the optimal $\sigma$.

As this example demonstrates, it is much more efficient to perform
a simulation at large noise levels. One can quickly try very many
moves even if most of them get rejected instead of just a few ones
where the energy difference has been accurately computed. 
However, there are practical problems with using large $\sigma$ as will be
discussed next.

\section{Uncertain Energy and Variance }
\label{sigmasec}

Unfortunately there is a serious complication: the variance needed
in the noise penalty is also unknown. Both the change in energy and
its variance need to be estimated from the data. The variance in
general will depend on the particular transition: $(s \rightarrow
s')$; we cannot assume it is independent of the configuration of
the walk. Precise estimates of variance of the energy difference
are even more difficult to obtain than of energy difference itself
since the error is the second moment of the noise and will
fluctuate more. In Fig.(\ref{doublewell}) is shown the effect on the
double well example of using an estimate of the variance in the
penalty formula instead of the true variance. The systematic error
arises because the acceptance rate formula is a non-linear
function of the variance. We will see that we must add an
additional penalty for estimating the variance from the data.

Let us suppose we generate $n$ estimates of the change in action:
$\{ y_1, \ldots , y_n\}$ where each $y_k$ is assumed to be an
independent normal variate with mean and variance: 
\be \langle y_i \rangle = \Delta\ee 
\be \langle (y_i-\Delta)^2 \rangle = n\sigma^2.\ee 
Unbiased {\it
estimates} of $\Delta$ and $\sigma^2$ are: \be\delta = \frac
{\sum_{i=1}^n y_i}{n}\ee \begin{equation}\chi^2 =
\frac{\sum_{i=1}^n (y_i-\delta)^2}{n(n-1)}.
\end{equation} By construction $\langle \delta \rangle =\Delta$ and
$\langle \chi^2 \rangle=\sigma^2$. 

The joint probability distribution function of $\delta$ and
$\chi2$ is the product of a normal distribution for the mean and a
chi-squared distribution for the variance:
\be
P(\delta, \chi^2 ; \Delta , \sigma ) = P(\delta-\Delta , \sigma )
P_{n-1} (\chi^2 ; \sigma) \ee where $ P(\delta - \Delta , \sigma )
$is given in
 Eq.(\ref{normal})
and
\be
P_{n-1} (\chi^2; \sigma) = c_n \chi^{n-3}e^{-\mu\chi^2/\sigma^2}
\ee with $\mu=(n-1)/2$ and \be c_n= \frac{(\mu/\sigma^2
)^{\mu}}{\Gamma(\mu)}. \ee

The generalization from the previous section is straightforward.
The acceptance probability can only depend on the estimators
$\delta$ and $\chi^2$. The average acceptance probability is:
\be
A(\Delta,\sigma)= \int_{-\infty}^{\infty} d\delta \int_0^{\infty}
d \chi^2 P (\delta,\chi^2; \Delta , \sigma ) a(\delta,\chi^2). \ee
Detailed balance requires:
\be
\label{twodeq}
A(\Delta,\sigma)= \exp(-\Delta) A(-\Delta,\sigma)
\ee 
for all values of $\Delta$ and $\sigma \geq 0$. We have two parameters to
estimate and average over instead of one and a two dimensional
homogeneous integral equation for $a(\delta,\chi^2)$.

In the limit of enough independent evaluations we recover the one
parameter equation since $ \lim_{n \rightarrow \infty} P_{n-1}
(\chi^2) = \delta(\chi^2-\sigma^2)$ and the equations for
different $\sigma$'s decouple.

\subsubsection*{Asymptotic Solution}

We can do the same type of analysis at large $|\Delta|$ as we did
when $\sigma$ was known. A move is definitely uphill or downhill
if $\delta^2 \gg \chi^2$. Assume there exists a solution with
$A(\Delta, \sigma) =1$ for $\Delta \ll -\sigma$. Then
$A(\Delta,\sigma)=\exp(-\Delta)$ for $\Delta \gg \sigma$. Assume
this solution can be expanded in a power series in $\chi^2$,
$a(\delta, \chi^2) = \sum_{k=0}^{\infty} b_{k} \chi^{2k}
e^{-\delta} $. Explicitly performing the integrals we obtain: \be
\exp(-\sigma^2/2) =\sum_k c_n
b_k\Gamma(\mu+k)(\sigma^2/\mu)^{\mu+k}.\ee Matching terms in
powers of $\sigma^2$ we obtain $b_k$.  The expansion can be summed
to obtain a Bessel function:
\begin{equation}
a(\delta, \chi^2) = \Gamma(\mu) e^{-\delta}
\left[\frac{2}{\mu\chi^2}\right]^{(\mu-1)/2}J_{\mu-1}
(\chi\sqrt{2\mu}).
\end{equation}
This function is positive for $\chi^2 < n/4$. For larger values of
$\chi^2$ either the assumption of $A(\Delta,\sigma)=1$ is wrong or
no smooth solution exists.

Taking the logarithm of the power series expansion, we obtain a
convenient asymptotic form for the penalty in powers of $\eta
=\chi^2/n$:
\begin{equation}
u_B = \frac{\chi^2}{2} +\frac{\chi^4}{4(n+1)}
+\frac{\chi^6}{3(n+1)(n+3)} + \ldots \label{bess}
\end{equation}
The ``Bessel'' acceptance formula is: \be a_B(\delta,\chi^2,n) =
\min(1,\exp(-\delta -u_B ))\ee.

The first term $\chi^2/2$, is the penalty in the case where we
know the variance. The error in the error causes an additional
penalty equal, in lowest order, to $\chi^4/(4n)$. This asymptotic
form should only be used for small values of $\eta$ since the
expansion is not convergent for $\eta \geq 1/4$. In Fig. \ref{errors} we
show errors in the detailed balance ratio as a function of
$\Delta$ and $\sigma$ for $n=128$. It is seen that the errors are
small but rapidly increasing as a function of $\sigma$. We find
that the maximum relative error in the detailed balance ratio
approximately equal to $0.15 \eta^2$. Good MC work will have the
error less than 10$^{-3}$ requiring $\eta < 0.1$ Very accurate MC
work with errors of less that 10$^{-4}$ requires a ratio $\eta <
0.02$. This is a limitation on the noise level.

As an example, we have calculated the deviation of the energy from
its exact value for the double well potential.
The results for the relative error in the energy are shown in Figure 6 for 
several values of $n$ and $\sigma$.
As we expect, the error in the energy depends only on $\eta$ and is proportional
to $\eta^2$.   We also see that the estimates of limits on the noise
level given above are correct.
There is a dip at $n=64$ for $\eta \approx .5$,
beyond the region where the Bessel expansion is convergent.

Figure \ref{eff} shows the effect on the efficiency of the additional
noise penalty. While the effect on the even moments is small, the
efficiency of the first moment dramatically increases for noise
levels $\sigma > 2$, perhaps because rejections for large
dispersions of the energy differences cause difficulty in crossing
the barrier. The efficiency becomes more sensitive to $\sigma$.

We have not found an exact solution for Eq. (\ref{twodeq}).
From numerical searches it is clear that much more accurate
solutions exist than the asymptotic form. We have found such
piecewise exponential forms. 
But the Bessel formula is a practical way of achieving detailed
balance if one can generate enough independent normally
distributed data.

\section{Deviations from a normal distribution}

We have assumed that $\delta$ is normally distributed. In the case
the noise is independent of position but otherwise completely
general, we can perform the asymptotic analysis. Let us assume
that:\be A(\Delta)=\int d\delta P(\delta-\Delta) a(\delta) \ee and
that $A(\Delta)=1$
for sufficiently negative values of $\Delta$.
Then for large values of $\Delta$ the unique continuous solution
is: \be a(\delta)=\exp(-\delta-u).\ee The penalty $u$ has an
expansion in terms of the cumulants of $P(\delta)$: \be
u=\sum_{n=2,4,\ldots}^{\infty} \kappa_n/n!=
-\ln(\int_{-\infty}^{\infty}dx P(x)e^{-x} ).\label{cumulant}\ee
The odd cumulants vanish because $P(x)=P(-x)$. For the normal
distribution this reduces to Eq. (\ref{1psol}) and the penalty
form is exact. The contribution of higher order cumulants could be
either positive or negative leading to positive or negative penalties.

Eq. (\ref{cumulant}) illustrates a limitation of the penalty
method: one can not allow the energy difference to have a long
tail of large values. It is important that the energy difference
be bounded because a penalty can be defined only if $\lim_{x
\rightarrow\infty} e^x P(x) =0$ so the integral will exist.
Suppose the energy difference in Eq. (\ref{sumpot}) is a sum of an 
inverse power of the
distances to the other particles $\Delta = \sum_{j} r_j^{-m}$ and
that ${\bf r}$ is sampled uniformly. Then we find (in 3
dimensions) that at large $\delta$: $P(\delta) \propto
\delta^{-3/m}$. For any positive value of $m$ the higher order
cumulants  and the penalty will not exist even though the mean and
variance of $\delta$ exist under the weaker condition: $m <3/2$.
We must arrange things so that large deviations  of the energy
difference from the exact value are non-existent or exponentially
rare, perhaps by bounding the energy error.

\section{Comparison with other methods}
\subsection{Method of Kennedy, Kuti, and Bhanot}

Kennedy, Kuti and Bhanot\cite{kk,bhanot} (KKB) have introduced a
noisy MC algorithm for lattice gauge theory. We adapted that
method for the present application by using energy differences
with respect to an approximate potential, $w(s)$, that can be
determined quickly and exactly. Proposed moves are
``pre-rejected'' using $w(s)$ and then the more expensive
computation of an estimate of $v(s)$ is done. Let us suppose that
the deviation between these potentials
can be bounded: $ \max |\delta w(s)- \delta v(s)|
\leq \epsilon$ for some $\epsilon$. We determine an unbiased
estimate of the ratio $q$ needed in Eq. (\ref{qdef}) by using the power
series expansion: 
\be 
q(s \rightarrow s') = e^{-\delta } =\sum_{n=0}^{\infty}(-\delta )^n/n! 
\ee 
where $\delta (s\rightarrow s') = v(s')-w(s')-v(s)+w(s)$. With a predefined
probability we sample terms in the power series up to order $n$
and obtain an estimate of $q$;
this is a variant of the von Neumann-Ulam method 
We finally accept the move with probability 
\be a
=(1+q)/(2+\epsilon).
\ee 
For an appropriate choice of parameters, $a$ is in the range $0\leq a \leq 1$
most of the time.
The revised KKB method is given by the following pseudo-code:

\begin{tabbing}
Sample $s'$ from $T(s\rightarrow s')$  \\ If
\=($\exp\left[-w(s')+w(s)\right] < \text{prn})$ then \\
   \> reject move  \\
else   \\
   \> $q_0=t_0=1$ \\
   \> do \=$n=1,\infty$  \\
   \>    \> $p_n = \min(\gamma/n,1)$\\
   \>    \> if ($p_n <$ prn) exit loop \\
   \>    \>  sample $x_n= -v(s') + w(s') + v(s) - w(s)$    \\
   \>    \>  $t_n = t_{n-1} x_n/(n p_n) $ \\
   \>    \>  $q_n = q_{n-1} + t_n$ \\
   \> end do \\
if $\left[(1+q_n)/(2+\epsilon)>\text{prn} \right]$ then accept move
\\

\end{tabbing}

In this procedure $\gamma > 0$ is a parameter which controls the
number of terms sampled. 
For $\gamma \leq 1$ the average number of
evaluations of $x$ per step is $n_e(e^{\gamma}-1)$, where $n_e$ is
the acceptance ratio of the preliminary rejection step.
Each sample of $x$ must be uncorrelated with previous samples. 
As $\epsilon \rightarrow 0$, one recovers the Metropolis algorithm.
The sampling distribution, $\gamma$, and $\epsilon$ have to be fixed to
ensure that $a$ is in the interval $[0,1]$ almost all of the time.
Violations for which $a<0$ put a limit on the size of the noise
and the size of the sampling step, while $\epsilon$ can be made
arbitrarily large to remove violations where $a>1$. 
This will, however, affect the efficiency. 
A recent preprint\cite{lin}
proposed to solve the problem of violating the constraints on the
acceptance probabilities by introducing negative signs into the
estimators. We have not explored this possibility.

We made a comparison to the penalty method with the double well
potential, using $w(s) = a_2 s^4$ as the approximate potential. (It
confines the random walk but does not have the central barrier.)
For a violation level of $10^{-4}$, the maximum noise was $\sigma=0.4$.
This is a much smaller noise level than is optimal in the penalty method.
For this noise, a transition step of 0.45 was optimal. 
To optimize $\gamma$ and $\epsilon$, we first adjusted $\gamma$
until the half the desired number violations occurred for $a<0$.
Then we adjusted $\epsilon$ until the total number of violations
equaled $10^{-4}$. 
The errors in the first and second moments are
given in Table \ref{kkbdat},  along with the parameters used in the KKB
algorithm.
We find that the KKB
method is 2.3 times slower for the first moment and 3.5 times
slower for the second moment than the penalty method (run at the
same noise level, with the same transition step size and computing
the variance with $n=32$ points). 
This comparison was done assuming
$f$ is sufficiently small that we do not have to take into account
the multiple evaluations of the energy differences. Taking that
into account would raise the inefficiency of the KKB method by
another factor of 2.74, the average number of function
evaluations.
 
We also tested the KKB method with $w(s)=v(s)$ 
({\it i.e.}, the argument of the exponential was only noise).
The data for this case is also given in Table \ref{kkbdat}.
The maximum value of allowable noise was still $\sigma = 0.4$.  
For $\sigma<0.2$, the average number of function evaluations was
less than one, making the method more efficient than the penalty method, 
for a fixed noise level.   For the first moment, KKB was 3.4 times more
efficient for $\sigma=0.1$ and 1.3 times more efficient for $\sigma=0.2$.
However, if we consider optimizing $\sigma$ as in Sec. \ref{noiseopt},
the KKB method is less efficient than the penalty method.
To be efficient at large values of $f$,
larger values of $\sigma$ must be used, and there the KKB method is less
efficient.  At small values of $f$, the last term in Eq. (\ref{effeq})
dominates, and the lesser number of function evaluations yields no advantage
for the KKB method.

The KKB method requires taking enough samples to lower the noise
to an acceptable value. In contrast, the penalty method requires
taking enough samples to ensure the distribution is normal. 
Also, for this problem, 
the penalty method could have an even higher efficiency because it
could use larger sampling steps sizes
(the maximum KKB sampling step size depends on the quality of the approximate 
function, $w$).
The advantage of the KKB method is that it makes no
assumptions about the normalcy of the noise; the disadvantage is
that one cannot guarantee that $a$ is in the range $[0,1]$.
Knowledge that the noise is normally distributed allows one to use a much
more efficient method.

\subsection{Reweighting}
Another alternative noisy MC method is to combine the stochastic
evaluation of an exponential with the reweighting method. One can
perform a simulation with $w(s)$, generating a random walk
$\{s_i\}$. Then an exact average can be generated by reweighting:
\be 
\langle O \rangle = \frac{\sum_i O(s_i) Q_i}{\sum_i Q_i} 
\ee 
where $Q_i \exp(
-(v(s_i)-w(s_i))/k_BT)$. As discussed above an estimate of $Q_i$
can be generated with the von Neumann-Ulam procedure by
stochastically summing the power series expansion of the
exponential. In this case we do not care whether the exponential
is between 0 and 1, only its variance is important. The difficulty
is that the exponent of the weight increases linearly with the
size of the system, {\it i.e.} $\langle (v(s)-w(s))^2 \rangle
\propto N$. Hence the variance of $\langle O\rangle$ will increase exponentially
with the size of the system. This method is only appropriate for
small systems, but no assumptions are made about the distribution 
of $v(s)-w(s)$.
The advantage of including the noise in the random
walk rather than reweighting the visited states is that one works
with energy differences only and it is possible to make the
fluctuations of  differences independent of the size of the
system.

\section{Conclusion}

We have shown a small modification of the usual random walk method
by applying a penalty to the energy difference can compensate for
noise in the computation of the energy difference. If the noise is
normally distributed with a known variance, the compensation is
exact. If one estimates the variance from $n$ data points, we show
that it suffices to have $\chi^2 \le 0.1 n$ and apply an
additional penalty. On a double well potential we found that the
the optimal noise level is typically $k_B T \leq \sigma \leq
3k_BT.$

The penalty method utilizes the power of Monte Carlo: one can
choose the transition rules to obey detailed balance and to
optimize convergence and use only well-controlled approximations.
We can generalize to other noise distributions by using
numerical solutions to the detailed balance equations as we have
shown. We have adapted a method introduced by Kennedy {\it et al.}
\cite{kk} but found it to be much slower once the noise level becomes high.

We now plan to apply the algorithm to a serious application. As we
have shown, very large gains in efficiency are sometimes possible.
However, the problem remains of ensuring that the estimates of
the energy differences are statistically independent and normally
distributed.

Codes used in calculations reported here are available at:
[http://www.ncsa.uiuc.edu/Apps/CMP/index.html ]

\section*{Acknowledgments}
This work was initiated at the Aspen Center for
Physics and has been supported by the NSF grant DMR 98-02373 and computational
facilities at NCSA.
DMC acknowledges useful discussions with J. Kuti, A. Kennedy and A.
Mira.

\begin{figure} 
\caption{The optimal acceptance formula
computed using the linear programming method (solid line) and
using the penalty method (dotted line). Both are for $\sigma=1$.
The accuracy of the LP solution is better than 1 part in $10^{-5}$.}
\label{LP}
\end{figure}

\begin{figure} 
\caption{The logarithm of the acceptance ratio as a function of 
$\sigma^2$ for the double well potential. The dashed line is
proportional to  $\exp(-\sigma^2/2)$ and the dotted line to
$\exp(-\sigma^2/8)$ }
\label{accratio}
\end{figure}

\begin{figure} 
\caption{The density as computed using the Metropolis formula (dotted line),
the direct penalty (dashed line), 
and the Bessel penalty with $N=16$ (solid line).
In all cases the noise level was $\sigma=2$.}
\label{doublewell}
\end{figure}

\begin{figure} 
\caption{The relative in-efficiency of penalty MC as a function of $\sigma$
and the noise level, $f$. From bottom to top the values of $f$ are 0.5, 1, 2, 4.
The solid lines are assuming the noise is known, the dashed lines are using the
Bessel formula with $n=64$ independent evaluations. Fig. (4a) is the first
moment, (4b) is the second moment.}
\label{eff}
\end{figure}

\begin{figure} 
\caption{The log of the detailed balance ratio versus $\Delta$ using
the Bessel penalty in Eq. (32) with $n=128$ points to estimate the variance.
From top to bottom the curves are with noise levels of $\sigma = 2, 1.8,
1.6, 1.4,$ and $1.2$.  }
\label{errors}
\end{figure}

\begin{figure}
\caption{The relative error in the energy for a double well potential 
versus $\eta$ for several values of $n$.
The circles are for $n=8$, the diamonds are for $n=16$, 
the squares are for $n=32$, and the triangles are for $n=64$. 
The dashed line has a slope of two.
}
\label{enery_error}
\end{figure}

\newpage
\begin{table}\caption{Computed MC inefficiencies for the modified
KKB method and the penalty method. $n$ is the average number of function
evaluations per step. The Bessel penalty method uses 32 data points and a 
transition step of 0.45.}
\label{kkbdat}

\begin{tabular}{dddddddd}
        &  \multicolumn{2}{c}{$c(\sigma)$}&\multicolumn{2}{c}{KKB parameters} \\
$\sigma$  & $x$ & $x^2$ & $\gamma$ & $\epsilon$ & $n$  \\ \hline
            \multicolumn{6}{c}{KKB, $w(s) = a_2 s^4$}   \\ 
0.0   &  273 &  154    & 1.07 & 6.0 & 1.49  \\
0.1   &  270 &  152    & 1.1  & 6.0 & 1.52  \\
0.2   &  275 &  151    & 1.2  & 6.0 & 1.61  \\
0.3   &  282 &  155   & 1.2  & 6.0 & 1.82  \\
0.4   &  270 &  145    & 2.1  & 4.9 & 2.74  \\
\multicolumn{6}{c}{KKB, $w(s) = a_1 s^2 + a_2 s^4$} \\
0.1   &  212 &  92    & 0.20 & 2.0 & 0.20  \\
0.2   &  209 &  89    & 0.59 & 1.8 & 0.59  \\
0.3   &  221 &  98    & 0.75 & 2.3 & 1.02  \\
0.4   &  215 &  98    & 1.35 & 2.1 & 1.93  \\
\multicolumn{6}{c}{Penalty} \\
0.0   & 175 & 74    \\
0.1   & 174 & 78    \\
0.2   & 184 & 76    \\
0.3   & 183 & 76     \\
0.4   & 178 & 78    \\ 
\end{tabular}
\end{table}


\begin{thebibliography}{99}
\bibitem{footnote1} 
The nuclei do not have to be classical:
one can treat quantum effects with imaginary time path integrals.
For simplicity, we do not discuss that case.
\bibitem{footnote2}
If the sampling probability were also
to be only estimated, in particular if the logarithm of $T$ was
normally distributed, that would simply add to $\sigma$ and the
discussion carries through.
\bibitem{footnote3}
However simply
requiring that $A(\Delta)=1$ for $\Delta \ll -\sigma$ does not
uniquely specify $a(\delta)$. Numerically we have found other
solutions with these boundary conditions.
\bibitem{footnote4}
A possible reason would be that one does not have
to compute an exponential or random number with such an acceptance
formula.
\end{thebibliography}
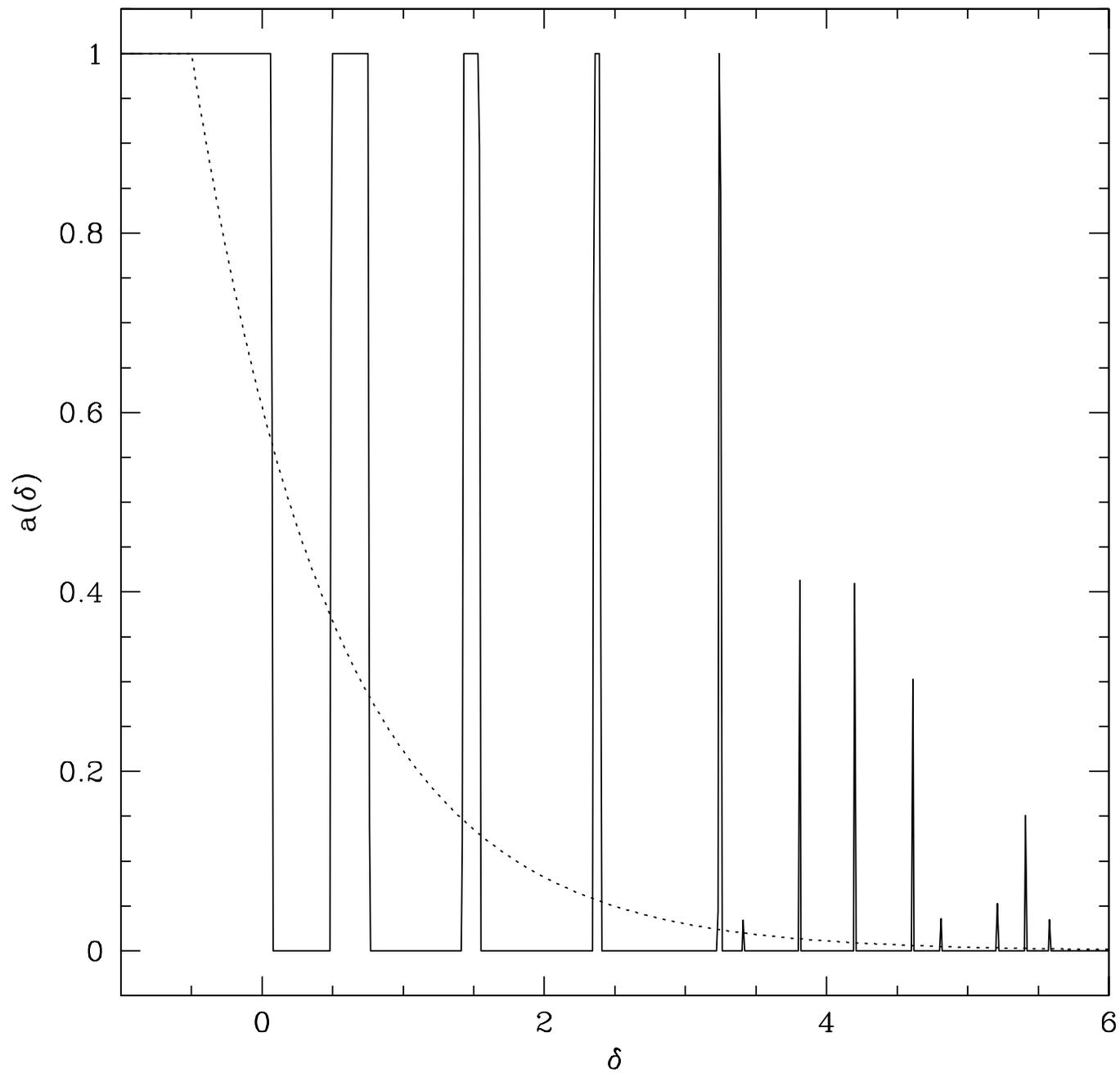
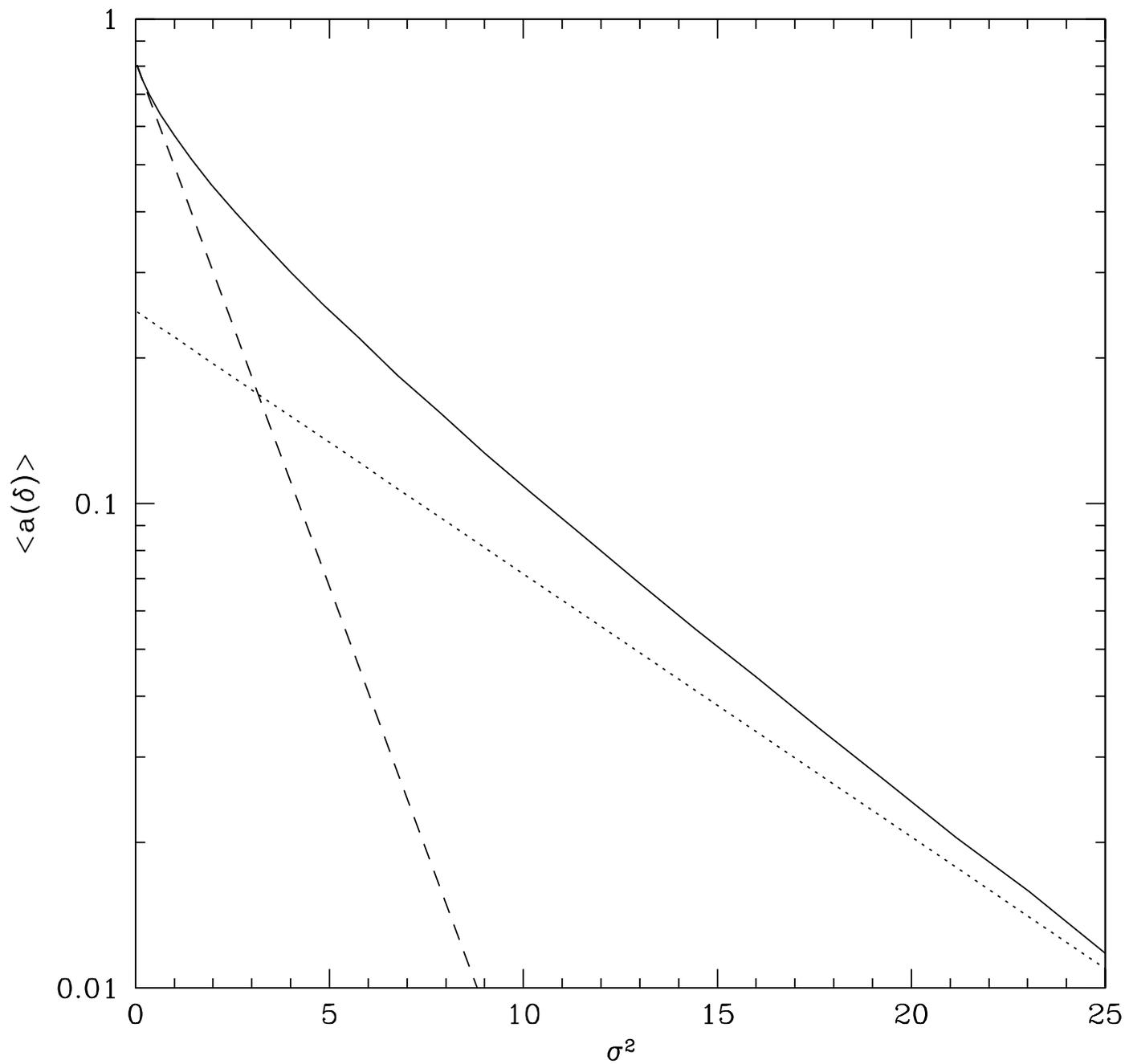
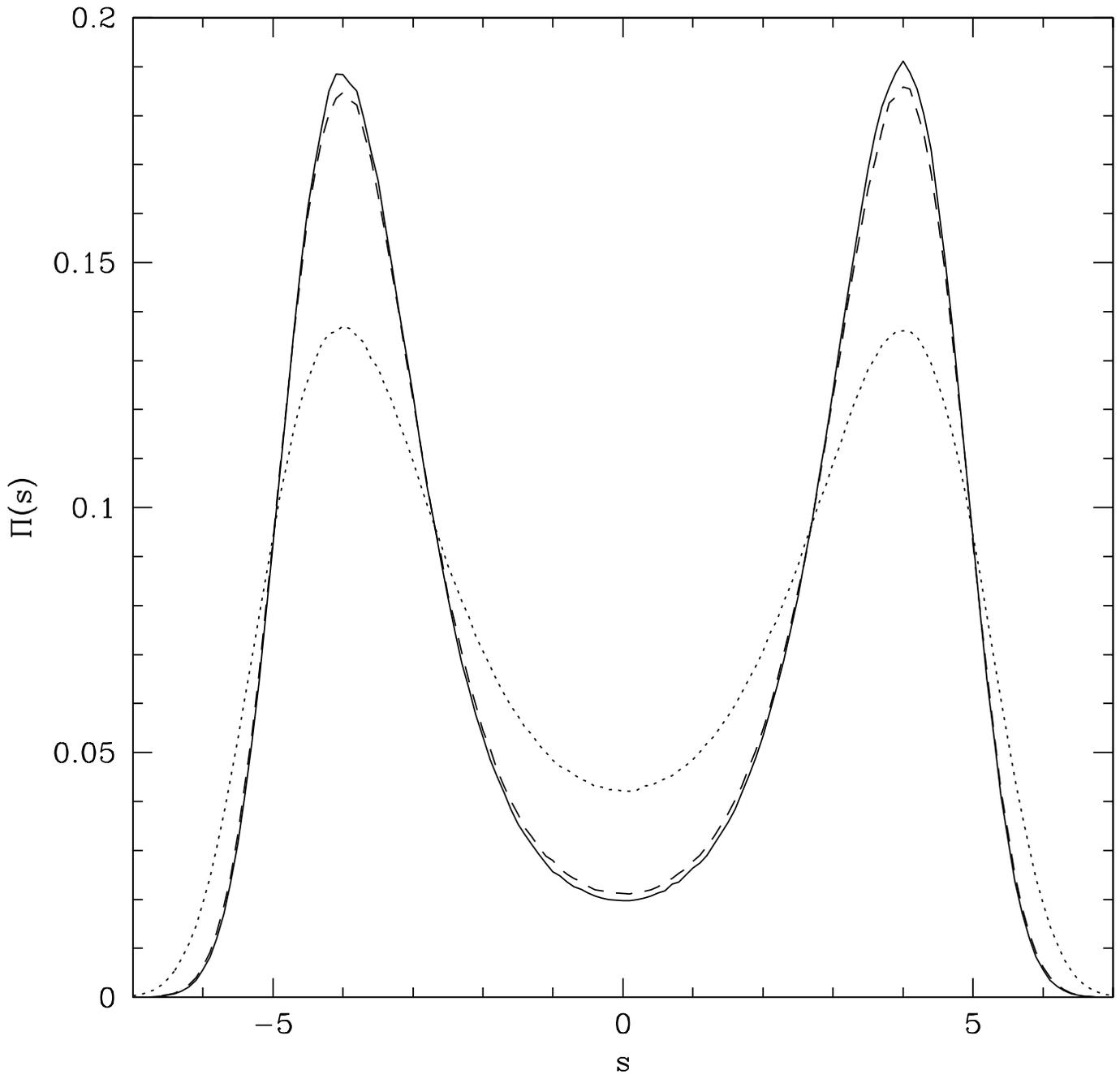
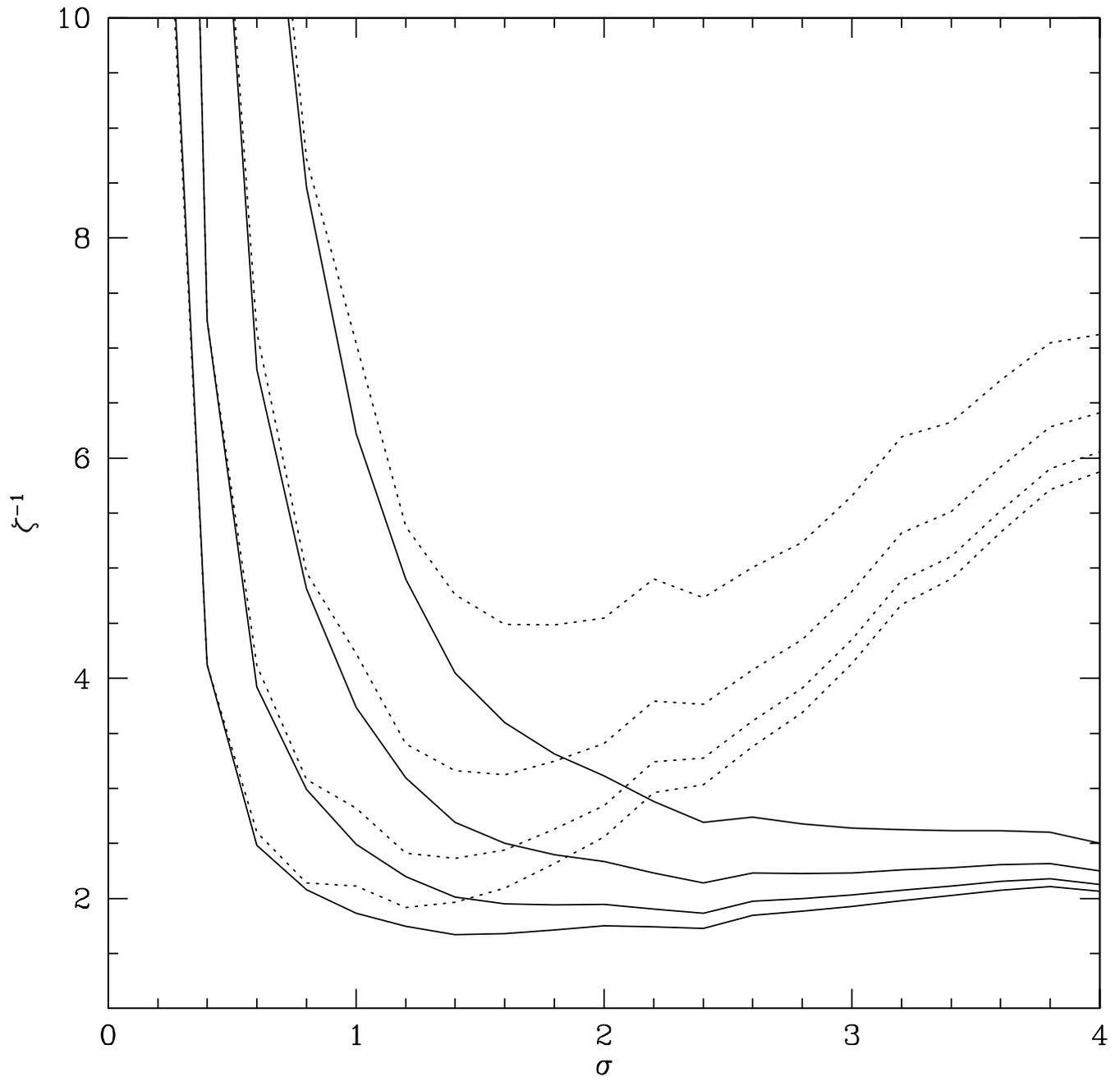
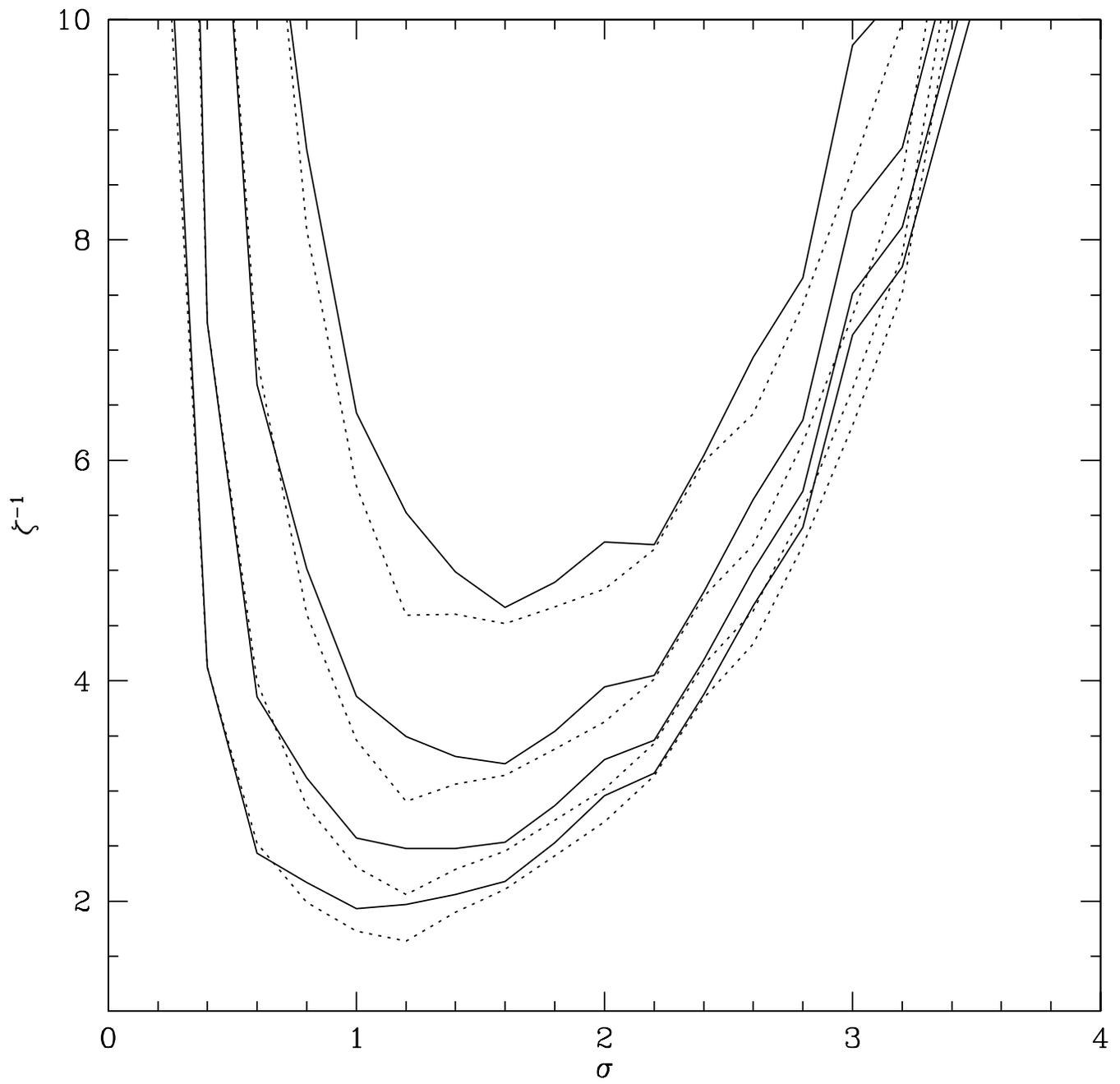
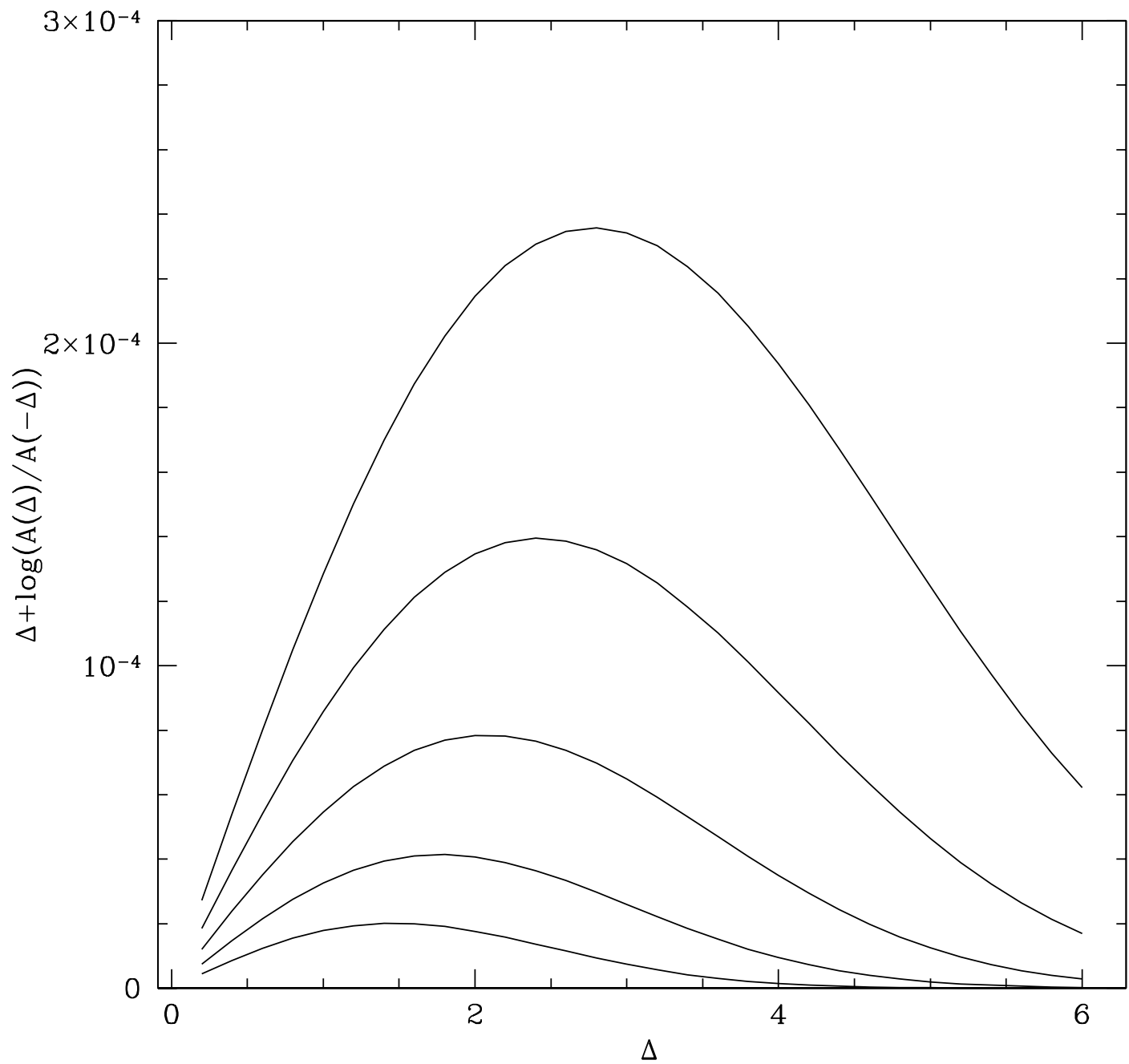
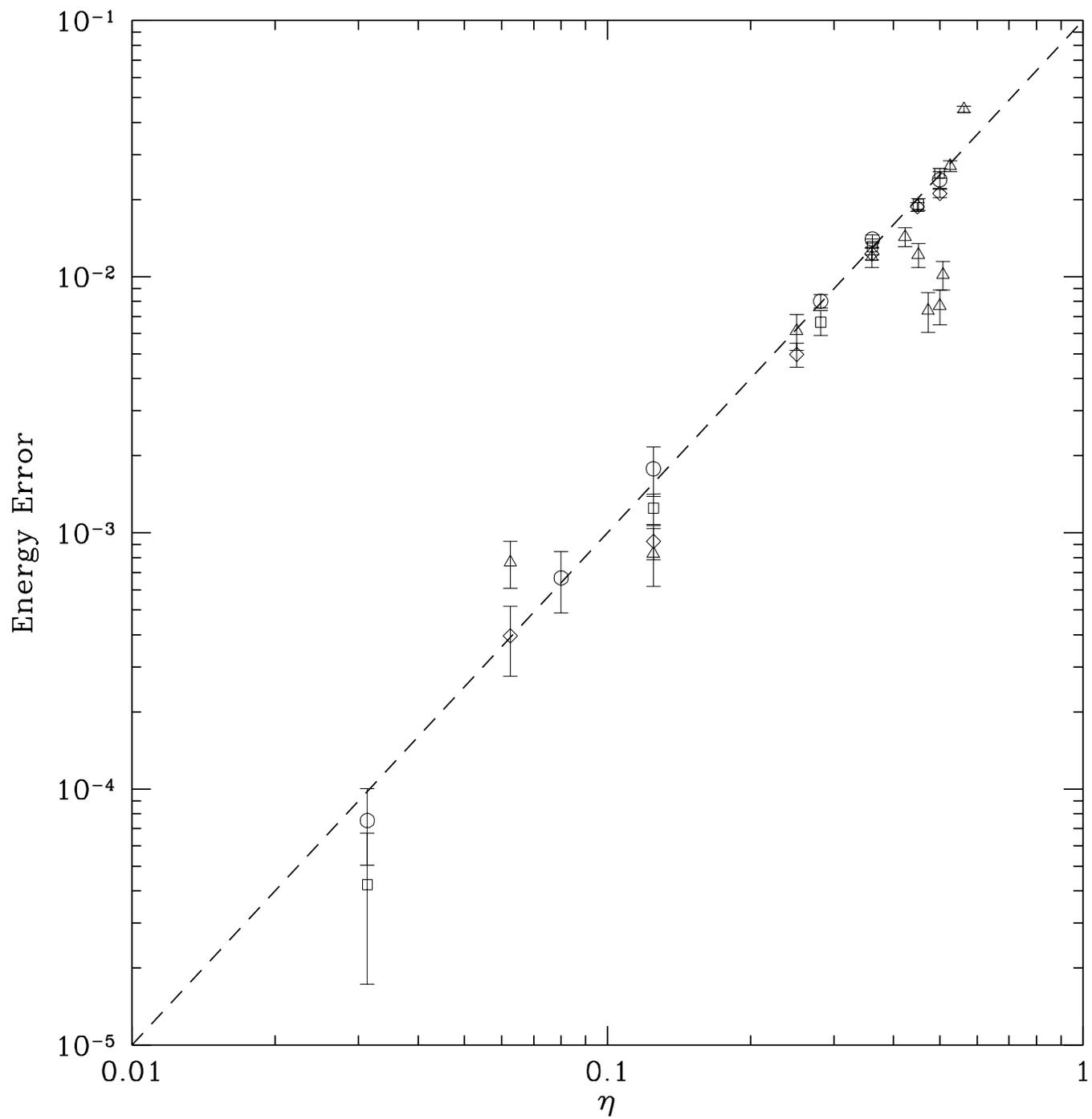
\end{document}